\documentclass[aps,prl,preprint,groupedaddress,showpacs]{revtex4-1}

\usepackage{graphicx}

\begin{document}

\title{Nonequilibrium phase transitions and violent relaxation in the Hamiltonian Mean Field model}
\author{T.~M.~Rocha Filho, M.~A.~Amato and A.~Figueiredo}
\affiliation{Instituto de F\'\i{}sica and International Center for Condensed Matter Physics\\ Universidade de
Bras\'\i{}lia, CP: 04455, 70919-970 - Bras\'\i{}lia, Brazil}
\date{}
\begin{abstract}
We discuss the nature of nonequilibrium phase transitions in the Hamiltonian Mean Field model using detailed numerical simulation of
the Vlasov equation and molecular dynamics. Starting from fixed magnetization waterbag initial distributions and varying the energy,
the states obtained after a violent relaxation undergoes a phase transition from magnetized to non-magnetized states when going from lower to higher energies.
The phase transitions are either first order or composed by a cascade of phase reentrances. This result is at variance with most previous
results in the literature mainly based in Lynden-Bell theory of violent relaxation. The latter is a rough approximation  and consequently
not suited for an accurate description of nonequilibrium phase transition in long range interacting systems.
\end{abstract}

\maketitle

%texto incluido MAA
The physics of long range interacting systems is an active topic of investigation due to the unusual and intriguing phenomenology
they present~\cite{maa1}. A pair potential interaction is considered long range if it scales at greater distances as $r^{-\alpha }$ with $\alpha <d$, where $r$ stands for the
inter-particle distance and $d$ the spatial dimension. This slow decaying interparticle potential is responsible for the coupling of distant
components of the system, a condition not encountered in short range systems.
A remarkable feature of these systems is that energy is non-additive and this opens up many nonintuitive phenomena, e.~g.\ in the microcanonical
ensemble it is possible to have negative specific heat and temperature jumps characterizing first order phase transition. In this context canonical and microcanonical
statistical ensembles can therefore be nonequivalent. Gravitational systems is another example that is largely studied~\cite{maa2a,maa2b,padma,maa2c} in the microcanonical ensemble,
and other systems not less important that encompass different areas of  physics, as plasmas~\cite{maa3}, wave-particle interactions~\cite{maa4} and
many others domains of application. A comprehensive review of the subject may be found in~\cite{maa1}.
These systems also present uncommon dynamical features.
Starting from an initial nonequilibrium configuration, these systems rapidly evolve by a violent relaxation to Quasi-Stationary States (QSS),
where they stay trapped for long lasting times scaled as an increasing function of the number of constituent particles,
and usually much longer than the time of observation that experimentalists are bound. Their structure was long ago recognized as non-Boltzmannian
states, and are now properly interpreted in terms of stable steady states of the Vlasov equation
and statistical equilibrium states in the sense of Lynden-Bell theory of violent relaxation~\cite{maa5a,maa8}.

Recently, a number of researchers studied nonequilibrium phase transitions in the Hamiltonian Mean Field (HMF) model~\cite{maa9}
in the context of Lynden-Bell theory~\cite{maa7a,maa7b,maa7c,maa7d,maa7e,maa7f,maa7g}.
They consider initial waterbag states with a given magnetization and looked for the final magnetization after the violent relaxation.
They then observed a phase transition from a magnetized to a non-magnetized QSS. Nevertheless the nature of such phase transitions and whether Lynden-Bell
theory correctly predicts them is still open to debate~\cite{maa7c,9b}.

In this paper we attempt to provide a more detailed description of the nature of nonequilibrium phase transitions of the HMF model,
and in particular, we pay attention to reentrant phases that seem to play an important and previously not fully acknowledged role.
We provide results from numerical simulations of the Vlasov Equation and Molecular Dynamics (MD). The HMF model is a system
of identical particles on a circle with unit mass and Hamiltonian:
\begin{equation}
H=\frac{1}{2}
\sum_{i=1}^N p_{i}^{2}+\frac{1}{2N}\sum_{i,j=1}^N
[1-cos(\theta _{i}-\theta _{j})],
\label{HMF}
\end{equation}
where $\theta _{i}$ is the angle that particle $i$ makes with a reference axis and $p_{i}$ stands for its conjugate momentum.
The $1/N$ factor in the potential energy corresponds to the Kac prescription to make the energy extensive and justify the validity of
the mean field approximation in the limit $N\rightarrow\infty$.
The relevant order parameter is the magnetization defined as:
\begin{equation}
M=\sqrt{M_x^2+M_y^2},
\label{magnet}
\end{equation}
where $M_x=(1/N)\sum_i\cos\theta_i$ and $M_y=(1/N)\sum_i\sin\theta_i$.

In the continuum limit the evolution of the single particle distribution function $f(\theta ,p,t)$ is governed by the Vlasov equation~\cite{9c,maa1}:
\begin{equation}
\frac{\partial f}{\partial t}+p\frac{\partial f}{\partial \theta }-\frac{dV[f]}{d\theta }\frac{\partial f}{\partial p}=0,
\label{vlasoveq}
\end{equation}
where $V[f]$ is is the interaction potential that depends self-consistently on $f(\theta ,p,t)$ and is given by
$V[f]\left( \theta \right) =1-M_{x}[f]\cos \left( \theta \right)-M_{y}[f]\sin \left( \theta \right)$, with
\begin{equation}
M_{x}[f]=\int d\theta\,dp\:f(\theta ,p,t)\cos \left( \theta \right),
\end{equation}
and 
\begin{equation}
M_{y}[f]=\int d\theta\,dp\:f(\theta ,p,t)\sin \left( \theta \right).
\end{equation}

In the foregoing discussion and following previous approaches~\cite{maa7a,maa7b,maa7c,maa7d,maa7e,maa7f,maa7g} we consider as initial state
a waterbag distribution, i.~e.\ $f(\theta,p,t=0)=1/2\Delta p\Delta\theta$ if $0<\theta<\Delta\theta$ and $|p|<\Delta p$, and
$f(\theta,p,t=0)=0$ otherwise. The initial magnetization and energy (per particle) are given by
$M=\left[\left(1-\cos\Delta\theta\right)^2+\left(\sin\Delta\theta\right)^2\right]^{1/2}/\Delta\theta$,
and $e=\Delta p^2/24+(1-M^2)/2$.

\begin{figure}[ptb]
\begin{center}
\scalebox{0.3}{{\includegraphics{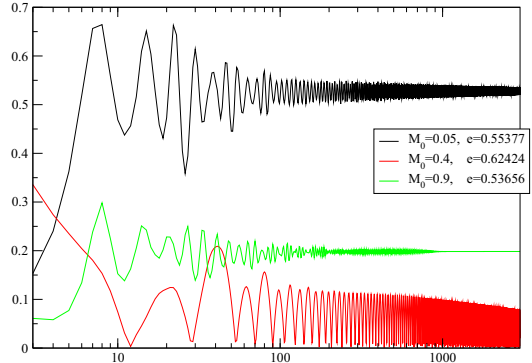}}}
\end{center}
\caption{Mono-Log graph of magnetization as a function of time for some initial magnetizations and energies per particle.}
\label{fig1}
\end{figure}

In order to discuss the out of equilibrium phases corresponding to the final state after a violent relaxation, it is
important to establish how long it takes for the system to settle down into a QSS or a possibly perpetually oscillating steady state~\cite{morita}.
Figure~\ref{fig1} shows the magnetization as a function of time for different initial values of magnetization $M_0$ and energy $e$.
It becomes clear that in many cases the QSS in only attained (or approached) for times of order $10^3$, at least one order of magnitude greater
than the total time used in some previous simulations on the same problem~\cite{antoniazzi,staniscia,buyl}.

Antoniazzi et al.~\cite{antoniazzi} compared the predictions from Lynden-Bell theory with N-body simulations at $E=0.69$ and 
obtained a reasonable agreement for $M_{0}<0.897.$ They also constructed a phase diagram
in the $\left( M_{0},e\right) $ plane and noticed that the system presents first and second order phase transitions
separated by a tricritical point. They proceed further and performed numerical simulations of the Vlasov equation and found 
reasonable agreement with N-body simulation and Lynden-Bell theory. Staniscia et al.~\cite{staniscia} in their calculations
confirmed the existence of reentrant phases as predicted by theory but show anyhow some
discrepancies and argued that this occurs due to incomplete relaxation during violent relaxation.

\begin{figure}[ptb]
\begin{center}
\scalebox{0.3}{{\includegraphics{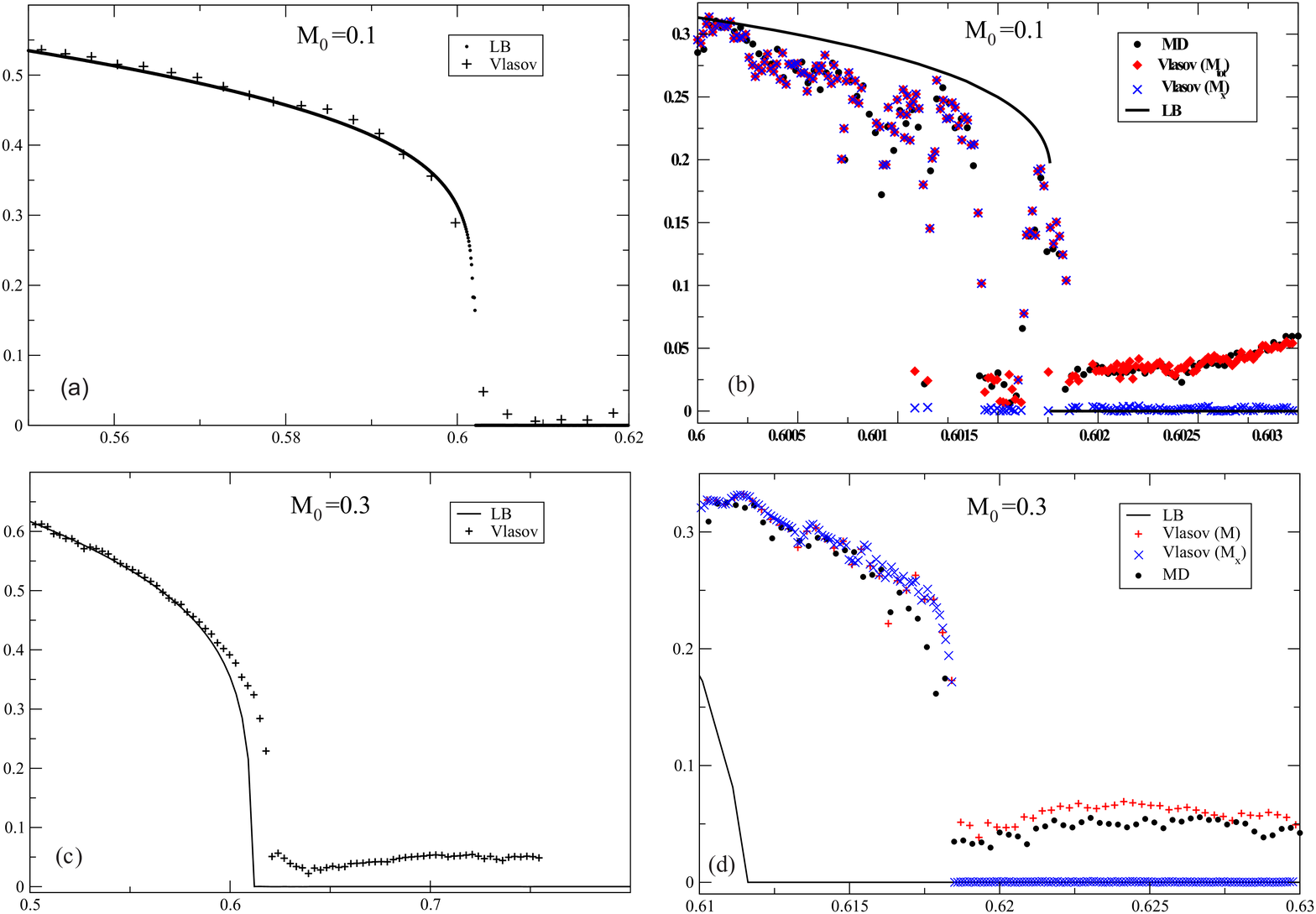}}}
\end{center}
\caption{QSS magnetization $M_f$ as a function of the energy per particle for initial magnetizations $M_0=0.1$ (panels a,b) and $M_0=0.3$ (panels (c,d).
All points from Vlasov solution were obtained using a numeric grid with $512\times512$ points and a time step $\Delta t=0.2$, total
integration time $t_f=3000.0$, and averaging from $t=2000.0$ up to $t=t_f$ except (b) that used a grid with $2048\times2048$
points, $t_f=1000.0$ and averaging from $t=800.0$ up to $t_f$. Molecular Dynamics simulations (MD) were performed with
$N=20,000,000$.}
\label{fig2}
\end{figure}

\begin{figure}[ptb]
\begin{center}
\scalebox{0.3}{{\includegraphics{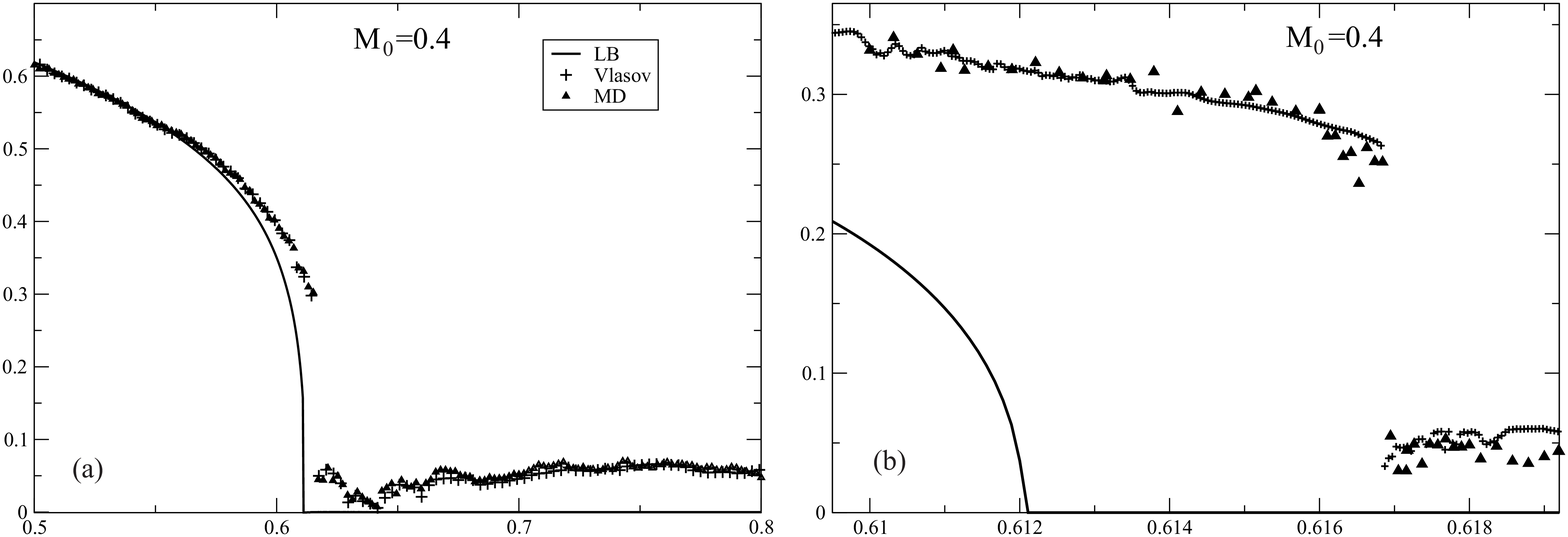}}}
\end{center}
\caption{QSS magnetization for $M_0=0.4$ computed from Lynden-Bell theory (LB), numeric solution of  Vlasov equation
and Molecular Dynamics (MD) with $N=20,000,000$, $t_f=3000.0$. The right panel shows in greater detail the discontinuity in the magnetization.}
\label{fig3}
\end{figure}

\begin{figure}[ptb]
\begin{center}
\scalebox{0.3}{{\includegraphics{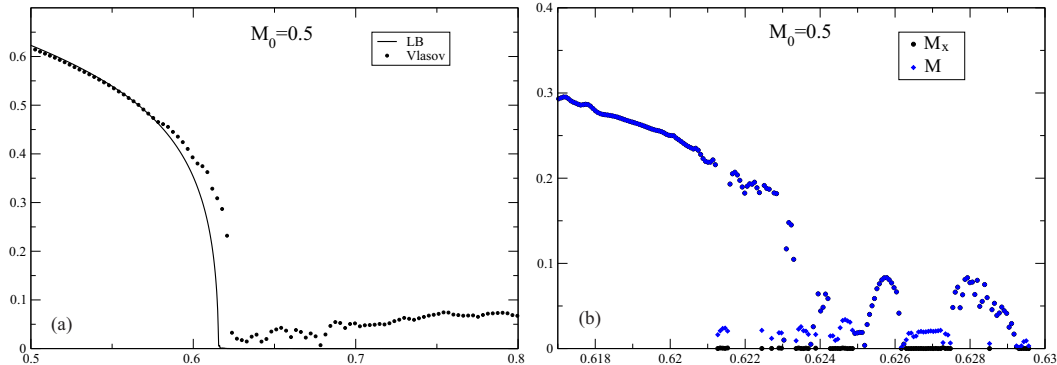}}}
\end{center}
\caption{Final magnetization as a function of energy for $M_0=0.5$ from the solution of Vlasov equation. The left panel (a) also shows the
prediction from Lynde-Bell theory (LB).}
\label{fig4}
\end{figure}

\begin{figure}[ptb]
\begin{center}
\scalebox{0.3}{{\includegraphics{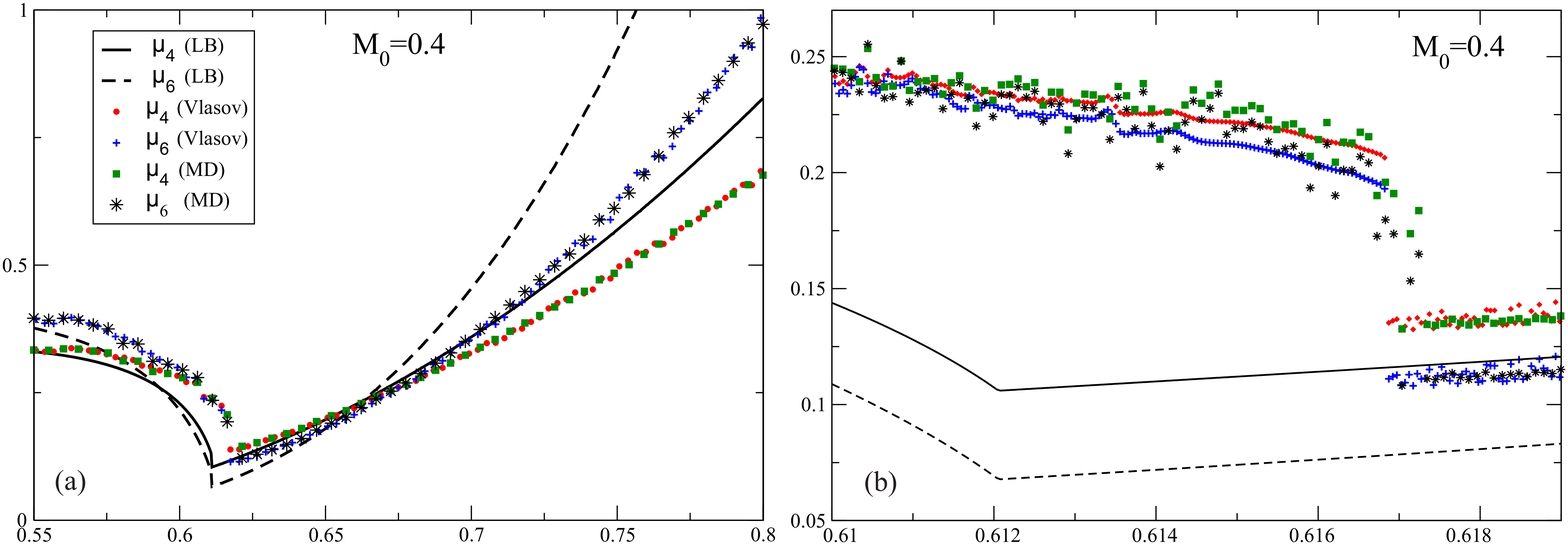}}}
\end{center}
\caption{Fourth and sixth moments of the velocity distribution function for $M_0=0.4$ as a function of energy computed
from Lynde-Bell theory (LB), the solution of the Vlasov equation and Molecular Dynamics (MD) with $N=20,000,000$.}
\label{fig5}
\end{figure}

Figures~\ref{fig2}--\ref{fig4} show the final magnetization as a function of energies for a few representative values of the initial magnetization
$M_0$ from the solution of the Vlasov equation, MD simulations and Lynden-Bell theory~\cite{maa11}.
Vlasov simulations were performed using a Vlasov integrator code in Ref.~\cite{rocha} with a numeric grid with
$512\times512$ points in the one particle phase space, total integration time $t_f=3000.0$ and averaging
from $t=2000.0$ to $t=3000.0$. For the more detailed graphics in figures~\ref{fig2}b, \ref{fig2}d, \ref{fig3}b, \ref{fig4}b and~\ref{fig5}b
we used a $2048\times2048$ grid with integration time $t_f=1000.0$ and
averaging from $t=800.0$ to $t_f$. The results from Lynden-Bell theory were obtained using the approach in~\cite{antoniazzi2}.
Figures~\ref{fig2}a and~\ref{fig2}c indicate that the transition is discontinuous in both cases predicted be from Lynde-Bell theory for
$M_0=0.1$ but not for $M_0=0.3$. This is even more clearly shown in Figs.~\ref{fig2}b and~\ref{fig2}d that show the region
near the phase transition using more simulation points and more accurate Vlasov simulations. For $M_0=0.1$ at least three reentrant phase transitions
are observed before the predicted (and observed) phase transition. The phase transition is more neatly observed for $M_0=0.3$,
where no mater the order parameter chosen ($M$ or $M_x$) the transition is clearly first order, from both MD and Vlasov equation solution.
In fact some discussion exists in the literature whether $M$ or $M_x$ should be used as an order parameter~\cite{maa7a}.
Here we argue that both choices lead to the same characterization of the order of the phase transitions.
For $M_0=0.4$ the discontinuity in the phase transition is even more evident as shown if Fig.~\ref{fig3}.
We note that for this particular value of magnetization the discontinuity in the phase transition was previously reported by Pakter and Levin~\cite{9b}.
They were also able to correctly predict the phase transition using a new ansatz for the distribution function based on dynamical properties
of the underlying Hamiltonian dynamics.
The situation gets even more interesting for $M_0=0.5$ in Fig.~\ref{fig4} where a close look around the phase transition reveals a cascade
of reentrant phases. As a consequence it is not clear how to asses the nature of the phase transition in this case.
For higher initial magnetizations the same analogous behavior is observed.

It is important to note that the critical energy of the phase transitions as predicted by Lynden-Bell theory is only an approximation, albeit a good one.
On the other hand, all previous
studies of nonequilibrium phase transitions in the HMF model have concentrate on the magnetization as an order parameter, which is obtained
from the spatial distribution function. It is interesting also to discuss what occurs with the velocity distribution function
along the same lines depicted previously. For that purpose we use the moments of the velocity distribution function given by
the average of powers of $v$ as $\mu_k\equiv\langle v^k\rangle$. Figure~\ref{fig5} shows the averaged moments $\mu_4$ and $\mu_6$ for $M_0=0.4$.
It is quite evident that Lynden-Bell theory gives reasonable results only for lower energies. The right panel of the same figures shows a discontinuity
in $\mu_4$ and $\mu_6$, a clear indication that the phase transition is indeed first order.

In this paper we have investigated phase transitions of QSS's using three different approaches: Lynden-Bell theory of violent relaxation,
numeric solutions of the Vlasov equation and molecular dynamics. Previous points in favor of Lynden-Bell theory
is that it gives a reasonable first approximation of the QSS's, and in this context, also allows to
predict out-of-equilibrium phase transitions, although it is also accepted the argument that the QSS's are, or can be, incomplete 
mixed stable states of the Vlasov equation~\cite{maa12}. The results presented here show unequivocally that the nature of phase
transitions is of first order are noticeable for different magnetizations, and reentrant phases are more common than previously noted,
as the cascade of phase reentrances observed for $M_0=0.1$ and $M_0=0.5$ clearly illustrates. The simulations also show 
that Lynden-Bell theory is not suitable to accurately predict these transitions. Molecular dynamics results are in very good agreement
with numeric solutions of the Vlasov equation. As a step forward we
have decided to calculate the moments (4th and 6th) of the velocity distribution and 
once more they diverge of those predicted by Lynden-Bell theory.  Although the latter, according to 
our calculations, is inadequate to explain nonequilibrium phase transitions in QSS's
it predicts with some accuracy the position of the phase transition, but not its order, and certainly not the phase
reentrances here reported. At lower energies it yields quite reasonable results for magnetization and moments of the velocity
distribution function, but strongly depart from the correct values at higher energies.
Therefore a detailed and accurate study of nonequilibrium phase transitions in long-range interacting systems cannot be based
on Lynden-Bell theory. Unfortunately a completely satisfactory theory for violent relaxation is still lacking, even though
some progress was obtained in Refs.~\cite{23} and~\cite{24}.

The authors would like to thank CNPq and CAPES (Brazil) for partial financial support.
TMRF would like to thank Y.~Levin and T.~Teles for fruitful discussions.


\begin{thebibliography}{99}
\bibitem{maa11} In the Lynden-Bell scenario for a two level initial state (waterbag state), the state after the violent relaxation
is obtained by maximizing the coarse-grained Lynden-Bell entropy $S_{LB}[F]=-\int dp\,d\theta\:[(f/f_0)\ln(f/f_0)+(1-f/f_0)\ln(1-f/f_0)]$
with the norm $\int dp\,d\theta\: f=1$ and energy
$e[f]=\int dp\,d\theta\: f(\theta ,p,t)p^{2}/2+(1-M[f]^2)/2$
constraints. The variational procedure provides the solution:
\begin{equation}
f_{QSS}\left( \theta ,p\right) =\frac{f_{0}}{1+\exp \left( \beta f_{0}\left( 
\frac{p^{2}}{2}+1-M_{x}[f_{QSS}]\cos \left( \theta \right)
-M_{y}[f_{QSS}]\sin \left( \theta \right) \right) \right) +\alpha },
\end{equation}
where $\alpha $ and $\beta $ are Lagrange multipliers.
\bibitem{maa12} Ergodicity is a topic that will be addressed in a future publication.
\end{thebibliography}
\end{document}